\documentclass[a4paper,11pt]{article}
\usepackage{cite}

\usepackage{jinstpub} 

\title{\boldmath CEE inner TOF prototype design and preliminary test results }

\author[a,b]{X.Wang}
\author[a,b,1]{D.Hu,\note{Corresponding author.}}
\author[a,b,1]{M.Shao,\note{Corresponding author.}}
\author[a,b]{L.Zhao}
\author[a,b]{Y.Sun}
\author[a,b]{J.Lu}
\author[a,b]{H.Xu}
\author[a,b]{Y.Zhou}

\affiliation[a]{State Key Laboratory of Particle Detection and Electronics, University of Science and Technology of China, 96 Jinzhai Road, Hefei 230026, China}
\affiliation[b]{Department of Modern Physics, University of Science and Technology of China (USTC), 96 Jinzhai Road, Hefei 230026, China}

\emailAdd{hurongr@ustc.edu.cn}

\abstract{The Cooling Storage Ring (CSR) External-target Experiment(CEE) is the first multi-purpose nuclear physics experimental device to operate in the GeV energy range at the Heavy-Ion Research Facility(HIRFL-CSR) in Lanzhou, China. The primary goals of the CEE are to study the bulk properties of dense matter and to understand the quantum chromo-dynamic (QCD) phase diagram by measuring the charged particles produced in heavy-ion collisions at the target region with large acceptance. An inner time of flight (iTOF) system has been proposed to measure the multiplicity, angular distribution, and time information of the charged particles. Herein, we introduce the performance requirements of iTOF according to calculations and GEANT4 simulations. The proposed system is characterized by high granularity and time performance, hence, the conceptual design of the iTOF wall adopts high granularity Multi-gap Resistive Plate Chambers (MRPC) with a time resolution around 30~\si{ps}. To evaluate the MRPC design, the cosmic ray test was performed. A timing resolution better than 28~\si{\pico\second} and an efficiency better than 98\% have been achieved for MIPs, as interpreted by the cosmic ray GEANT4 simulation of time jitter components.}


\begin{document}
\maketitle
\flushbottom

\section{Introduction}
\begin{figure}[htbp]
\centering
\includegraphics[width=.7\textwidth]{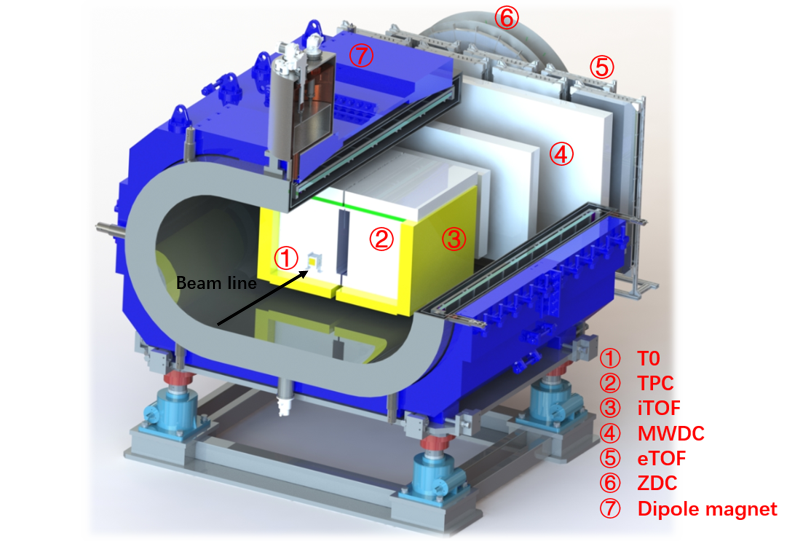}
\caption{ Schematic layout of the CEE spectrometer. }
\label{fig:1}
\end{figure}

Quantum chromodynamics(QCD)is a modern theory describing the strong interactions that govern more than 90\% of the visible matter in nature. Although nuclear physicists have significantly advanced QCD in recent years, many unknowns remain, especially the phase structure of strongly interacting matter under the QCD degrees of freedom. For example, we do not yet understand the phase transition of matter from the hadron degree of freedom that of quark-gluon, and the possible phase transition critical point~\cite{braun2007quest}. Moreover, We know little about the equation  describing the state of strongly interacting matter under various temperature and density conditions~\cite{xiao2014probing}. These are the central scientific issues involved in heavy-ion collision experiments. To study these issues and have a deeper understanding of QCD, various heavy-ion collision experiments are planned or underway, such as LHC-ALICE~\cite{ackermann2003star}, RICH-STAR~\cite{aamodt2008alice}, FIAR-CBM~\cite{friese2006cbm}, and NICA-MPD~\cite{toneev2007nica}. China is actively participating in the related research. In China, the HIRFL-CSR~\cite{xia2002heavy} and the high intensity heavy ion accelerator facility(HIAF)~\cite{yang2013high} under construction can provide beams of different nuclei types in the \si{GeV} energy region, which are important to study of high baryon density QCD matter. The CEE is a spectrometer focusing on charged final-state particle measurements running on HIRFL-CSR, and it will continue its operation, with possible upgrades, at HIAF\cite{lu2017conceptual}.
The CEE spectrometer works in a fixed target mode. It is designed to achieve high precision measurement of charged final-state particles with large detection acceptance. The conceptual design diagram of the CEE is shown in Fig~\ref{fig:1}. It comprises a set of sub-detectors, including beam monitor, T0 detector~\cite{hu2017t0}, time projection chamber (TPC)~\cite{li2016simulation}, inner TOF (iTOF), multi-wire drift chamber (MWDC)~\cite{sun2018drift}, external TOF (eTOF), and zero-degree calorimeter (ZDC)~\cite{zhu2021prototype}. It also has a large super-conducting dipole magnet. Two sets of measurement and identification systems for charged particles were designed for the CEE spectrometer. For high rapidity, or small angle region(0$\degree$ - 25$\degree$), the charged particles are identified by eTOF combined with MWDC; for medium rapidity, or large angle region(25$\degree$ - 107$\degree$),charged particles are identified by iTOF combined with TPC. In addition, the start times of eTOF and iTOF are provided by T0 detector. Here the TPC adopts a double-sensitive-volume design, with an outer envelope size of 1500~\si{mm}(X,horizontal)$\times$  920~\si{mm}(Y,vertical) $\times$  1200~\si{mm}(Z) (Z is along the beamline direction). There is a 130~\si{mm} wide gap in the TPC,the target is located at Z = -350~\si{mm} from the center of the TPC. To maximize the acceptance and detection efficiency, the iTOF system is divided into four regions close to the TPC in the conceptual design.

\section{iTOF system design}
\begin{figure}[htbp]
\begin{minipage}[t]{0.5\linewidth}
\centering
\includegraphics*[width=1\textwidth]{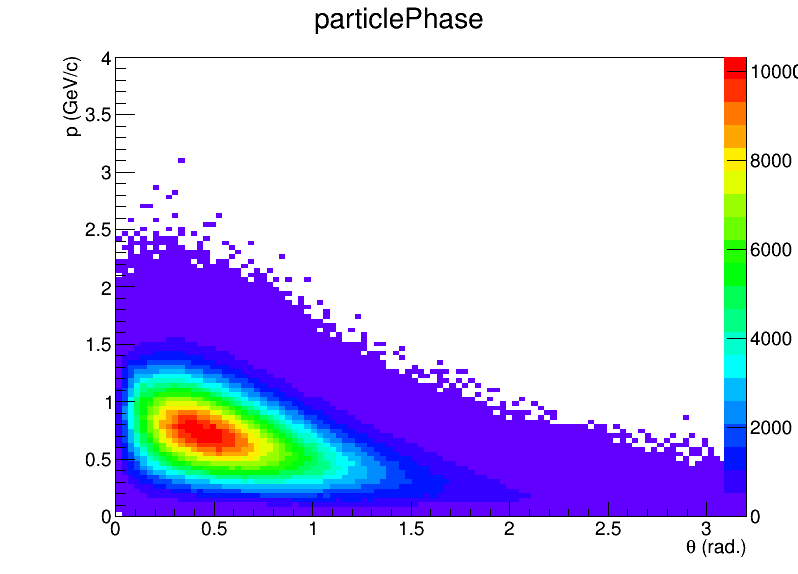}
\centerline{(a)}
\end{minipage}
\hfill
\begin{minipage}[t]{0.5\linewidth}
\centering
\includegraphics*[width=1\textwidth]{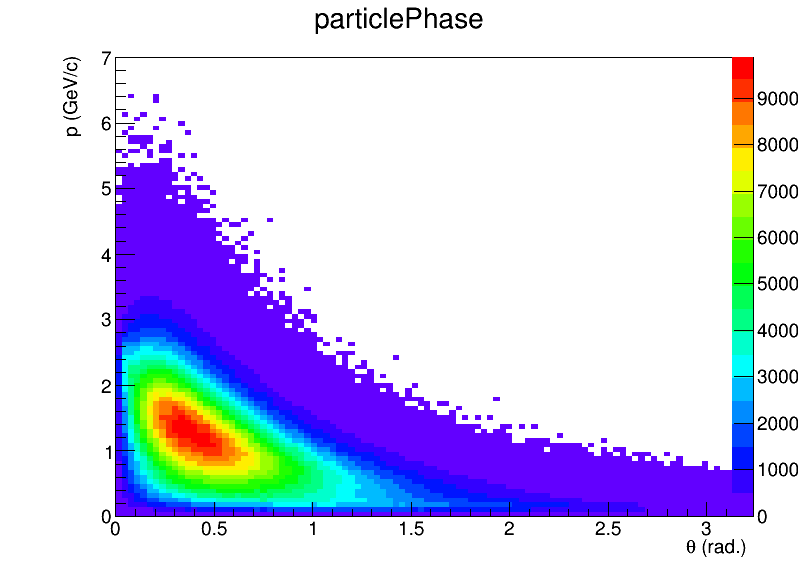}
\centerline{(b)}
\end{minipage}
\caption{  Phase space distribution of final-state charged particles of 500~\si{MeV/u} UU centering collision (left) and 1.5~\si{GeV/u} UU centering collision (right).}
\label{fig:2_1}
\end{figure}

The TOF system is critical for the identification of charged particles in the GeV energy region. An important characteristic of a TOF system is the time performance. The time resolution requirements should be calculated according to the category and momentum range of the particles to be identified, combined with the flight distance of the particles. In HIRFL-CSR, the beam energy is $\lessapprox$ 1~\si{GeV/u}. Fig~\ref{fig:2_1}(a) shows the phase space distribution of charged particles in the final-state of 500~\si{MeV/u} UU centering collision by UrQMD ~\cite{urqmd} generator,with the x-axis and y-axis represeting the particles' emission polar angle and momentum,respectively. Fig~\ref{fig:2_1}(a) also shows that most of the particles are concentrated arround 0.5 rad. degree, and the momentum of particles in the iTOF coverage is typically lower than 1.5~\si{GeV/c}; therefore, the requirement for PID is not high. However,for HIAF,as the beam energy is increased to a few GeV/u, the momentum of the final-state particles is also significantly increased. It is important to consider the time performance requirement of iTOF for high beam energies. Fig~\ref{fig:2_1}(b) shows the phase space distribution of charged particles in the final state of 1.5~\si{GeV/u} UU centering collision. The momentum of particles in the iTOF coverage increase to 2 - 3~\si{GeV/c}. Thus, higher requirements are placed on the performance of iTOF.

\begin{figure}[htbp]
\begin{minipage}[t]{0.5\linewidth}
\centering
\includegraphics*[width=.8\textwidth]{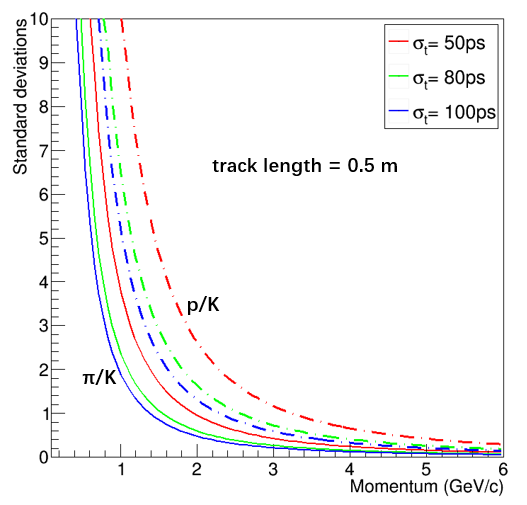}
\centerline{(a)}
\end{minipage}
\hfill
\begin{minipage}[t]{0.5\linewidth}
\centering
\includegraphics*[width=.8\textwidth]{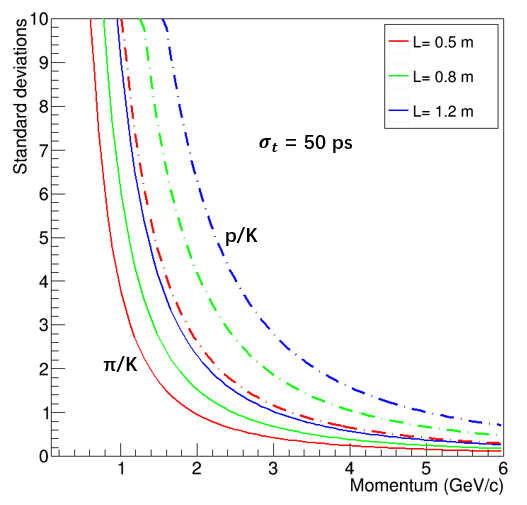}
\centerline{(b)}
\end{minipage}
\caption{ Variation of $\pi$/K and K/p separation power with momentum at 0.5~\si{m} flight distance (momentum dispersion is not considered).}
\label{fig:2_2}
\end{figure}

The goal of iTOF is to achieve the $\pi$/K/p separation of 3$\sigma$ for >95\% charged particles produced in heavy-ion collision at CEE. The separation power of particle identification can be defined as~\cite{lippmann2012particle}:
\begin{equation}
n_{\sigma}=\frac{|T_1-T_2|}{~\sigma}=   |m_1^{2}-m_2^{2}|\frac{L}{2cp^{2}~\sigma}
\end{equation}
Here, L is particle flight distance, p is particle momentum, $m_1$ and $m_2$ represent two different particle masses, and $\sigma$ is the time resolution. Fig~\ref{fig:2_2} shows the variation of $\pi$/K and K/p separation power with momentum at different flight distances and time resolutions.In this case, the time resolution includes the time jitter provided by the T0 detector. Thus,as seen in Fig~\ref{fig:2_2},for a overall time resolution of 50~\si{ps} in the condition of 0.5~\si{m} flight distance,the 3$\sigma$ separation power gives an upper momentum of 1.1~\si{GeV/c} for $\pi$/K separation, and 1.8~\si{GeV/c} for K/p separation. For a longer flight distance, such as 1.2~\si{m}, the 3$\sigma$ separation power extends to 1.8~\si{ GeV/c} and 2.8~\si{GeV/c} for $\pi$/K and K/p identification respectively, fulfilling the PID needs even at HIAF. A system time resolution of 50~\si{ps} means that the iTOF needs to achieve a time accuracy of $\sim$30~\si{ps}. 

\begin{figure}[htbp]
\centering
\includegraphics[width=1\textwidth]{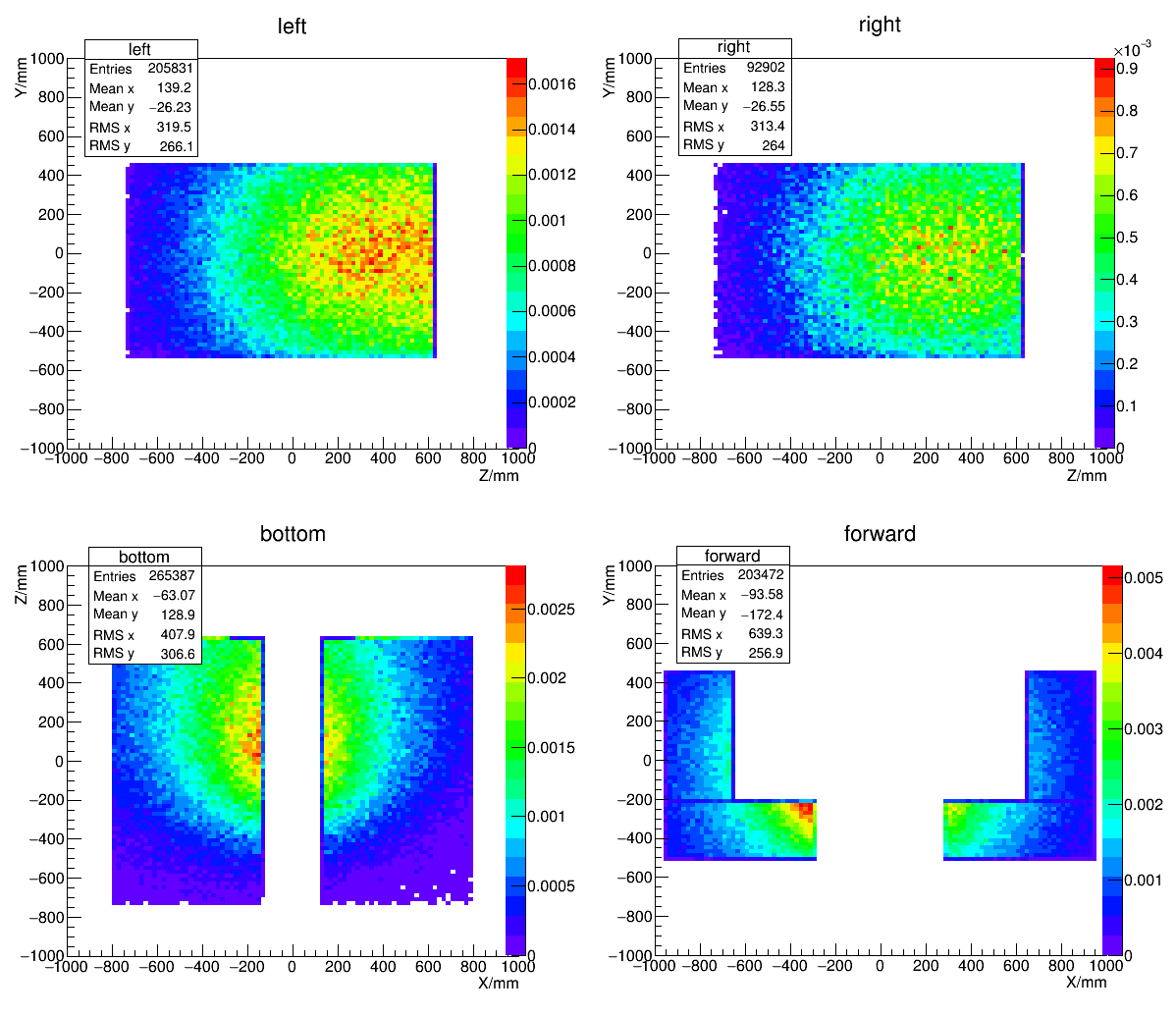}
\caption{ Track density on iTOF wall for 500~\si{MeV/u} U-U collision, the unit is per event per square centimeter.
 }
\label{fig:2_3}
\end{figure}

\begin{figure}[htbp]
\begin{minipage}[t]{0.5\linewidth}
\centering
\includegraphics*[width=1\textwidth,height=.8\textwidth]{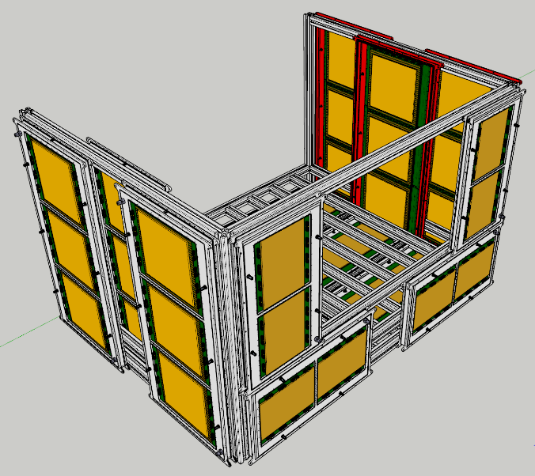}
\centerline{(a)}
\end{minipage}
\hfill
\begin{minipage}[t]{0.5\linewidth}
\centering
\includegraphics*[width=1.2\textwidth,height=.85\textwidth]{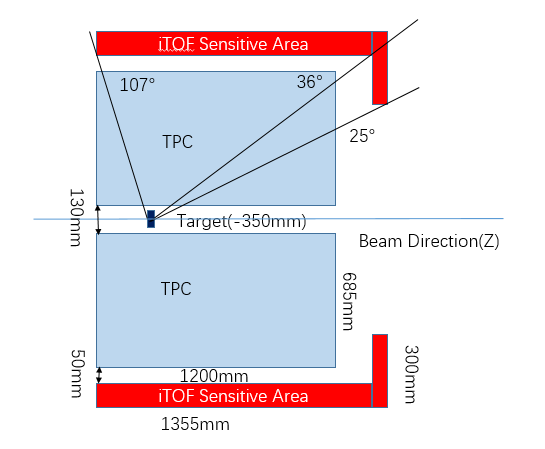}
\centerline{(b)}
\end{minipage}
\caption{ Layout of the iTOF system.}
\label{fig:2_4}
\end{figure}

Since the expected beam intensity of the CEE is $10^{6}/s$, and the reaction rate of the target is 1\%,it yields a total reaction rate of 10~\si{kHz}. Considering the high multiplicity of heavy-ion collision experiments, a high particle flux on the iTOF wall is foreseen. To ensure an occupancy of no more than 15\%, the granularity of iTOF should be high enough, to produce a total readout channel of $\sim$1500. We built a GEANT4~\cite{agostinelli2003geant4} simulation frame combined with 500~\si{MeV/u} UU collision events source generated by an IQMD~\cite{lv2014nuclear} generator to calculate the hit density of iTOF particles. In the simulation, four iTOF walls were placed close to the TPC sides, including the left side, right side, bottom side, and forward side along the beam direction. The left and right TOF walls have the same sensitive volume of 1355~\si{mm}(W)$\times$ 1000~\si{mm}(H)$\times$100~\si{mm}(T),and the shortest distance to TPC surface is 5~\si{cm}. The bottom TOF wall is divided into two equal parts, each with a sensitive volume of 1355~\si{mm}(W)$\times$ 670~\si{mm}(H)$\times$100~\si{mm}(T),and the shortest distance to TPC surface is also 5~\si{cm}. The forward TOF wall only covers part of the TPC surface, comprising four modules with a sensitive volume of 670~\si{mm}(W)$\times$ 300~\si{mm}(H)$\times$75~\si{mm}(T). Both TPC and iTOF walls were placed in a uniform magnetic field of 0.5~\si{T}. The simulation results are shown in Fig~\ref{fig:2_3}: the upper left and right graphs represent the hit density distribution on the left and right TOF walls; the lower left and right graphs represent the hit density distribution on the bottom and forward TOF walls. The simulation results illustrate that the hit density on the left and right iTOF walls is less than 0.0016~\si{hit/cm^2}, and the hit density of the right wall is lower than that of the left wall because of positively charged particles, which are produced much more abundantly than negatively charged particles.The hit density on the bottom iTOF wall is less than 0.0025~\si{hit/cm^2}. On the front iTOF wall,the hit density is relatively low except for the position close to the beam line, which is 0.005~\si{hit/cm^2}.The left, right, and bottom walls use detectors with a sensitive area of 45~\si{cm^2}, with an occupancy of 11.25\%.The forward wall employs a detector with a sensitive area of 25~\si{cm^2},with an occupancy rate of 12.5\%. It is clear that for such high granularity, the use of scintillators would incur huge costs; therefore the CEE collaboration has opted to use MRPC since this detector technique effectively balances time resolution, efficiency, and cost requirements.
Based on the above simulation and analysis,the structure of the iTOF system was determined, with the conceptual design shown in Fig~\ref{fig:2_4}. The total sensitive area of iTOF is $\sim$6\si{m^2}, covering the polar angle range from 25$\degree$ to 107$\degree$. The iTOF walls comprise 16 super modules(SMs) and a total of 38 MRPC detectors. Each SM is composed of two or three MRPC detectors according to the different regions. To achieve the time resolution requirements of PID, the ultra-high time resolution (\textless 30~\si{\pico\second} ) MRPC prototype with high precision readout electronics is adopted. In the following section, we describe the methods used to design and test the MRPC prototype, including the configuration of MRPC structure,readout electronics, and cosmic ray testing results.

\section{R$\&$D of iTOF-MRPC}
\subsection{Structural design}
The MRPC is a gaseous detector with excellent time resolution and high detection efficiency, invented in 1996 by the LAA project for the ALICE experiment~\cite{zichichi1987laa}.Typically,the time resolution of MRPC is approximately 50~\si{\pico\second},such as the MRPC3b for CBM TOF~\cite{hu2019beam}. The intrinsic time resolution of a single gap RPC is ~\cite{riegler2004physics}:
\begin{equation}
\sigma_t=\frac{1.28}{(\alpha-\eta)v}
\end{equation}
where $\alpha-\eta$ is the effective Townsend coefficient and v is the electron drift velocity in the gas gap.Based on this equation,the intrinsic time resolution of an MRPC depends only on the effective Townsend coefficient and the drift velocity, both increasing as the electric field E gets stronger. The movement of the electrons from an avalanche in the gas gap is rapid and generates the fast signal. The ratio of the fast signal to the total signal is $1/\alpha$D~\cite{akindinov2004space}, where $\alpha$ is the first Townsend coefficient and D is the gap width. According to these basic ideas, designing a high time resolution MRPC requires the electrons in the gas gap to drift rapidly,and the ratio of fast signal reasonably high. These two aspects can be fulfilled by increasing E, which in turn increases $\alpha$ and $v$ (thus improving $\sigma_t$), and decreasing D (which keeps fast signal ratio high). Therefore, building such a MRPC with time resolution better than 30~\si{\pico\second} while maintaining an efficiency close to 100\% is theoretically achievable. The main methodology to improve the MRPC's time resolution is to reduce the thickness of gas gaps and increase the number of gas gaps. This increases the ratio of the fast signal while reducing the time wiggle caused by the electron drift. At the same time, more gas gaps can also improve the detection efficiency.

\begin{figure}[htbp]
\begin{minipage}[t]{0.5\linewidth}
\centering
\includegraphics*[width=1\textwidth]{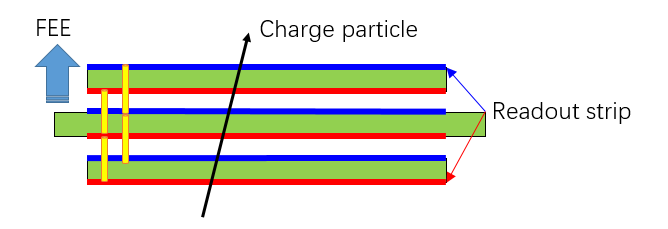}
\centerline{(a)}
\end{minipage}
\hfill
\begin{minipage}[t]{0.5\linewidth}
\centering
\includegraphics*[width=1\textwidth]{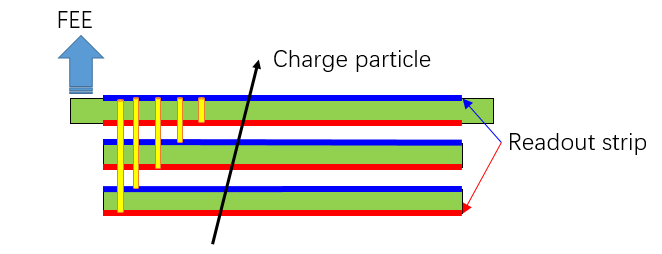}
\centerline{(b)}
\end{minipage}
\caption{ Schematic diagram of different MRPC readout structures.}
\label{fig:3_1}
\end{figure}

In addition, to ensure that a high resolution MRPC can be designed, the signal integrity of the detector must be considered. Signal integrity refers to the signal shape remaining as constant as possible during the transfer of the signal from the readout strip to the electronics. In the perfect case, the signal is output from the end of the readout strip and directly enters the front-end electronics (FEE) for discrimination and amplification. However, due to the MRPC structure, it is challenging for the FEE to be integrated with the readout board. In the general readout design, signals are induced from different layers of readout boards that are aggregated to the same readout board through signal pins, and then transmitted to the FEE through transmission lines and connectors. Every part of the transmission process affects the leading edge of the signal. In addition, for multi-layer MRPC, the signal is drawn from the middle PCB traditionally, as shown as Fig~\ref{fig:3_1}(a). Since the time of particles arriving at the different layers of the MRPC are different, the signals induced by each PCB also have time differences. Generally, it can be corrected by traces on the PCB. However, the longer the traces, the greater is the impact on the signal. To minimize this extra trace, the iTOF MRPC signal is led out from the bottom readout PCB (the readout PCB away from target during experiment) as shown as Fig~\ref{fig:3_1}(b). Moreover, the FEE is connected directly to the bottom readout PCB through a high-quality, high-bandwidth connector.

Another factor that affects the signal integrity is impedance matching, which is an important concept in signal transmission theory. For MRPC, the impedance mismatch brings two problems. One is that a reflected signal is generated. If the reflected signal is close in time to the original signal, it is superimposed on the original signal, distorting the signal shape. The other is causing the loss of the original signal charge, which is disadvantageous to the measurement of the signal. The MRPC signal is induced on the readout strip, passed through the signal pins, PCB traces, connectors, and finally transmitted to the FEE. The impedance of all parts should be matched as much as possible. The impedance of the MRPC  FEE is generally 110~\si{\Omega} differential,e.g... for NINO. But the impedance of MRPC itself is very difficult to be as high as this value. The MRPC can be simplified into a parallel structure of multiple parallel-plate capacitors, and its impedance is related to factors such as the distance between PCBs and the width/pitch of the readout strip. For example, the MRPC3b we designed for CBM TOF has 2$\times$5 gaps with a gap size of 230~\si{\mu m} and strip width of 7~\si{mm}~\cite{hu2019beam}. The impedance of this prototype is around 50~\si{\Omega} differential and matched to the preamplifier with a parallel resistor, which means more than half of the signal charge is dispersed. The iTOF MRPC plans to use a four-stack structure, allowing the impedance to be further reduced. Although the multi-stack structure increases the amount of induced signal charge, the fraction of signal flowing into the FEE will still be small, if the impedance matching is not properly dealt with. The finite-element method (FEM) calculation with ANSYS~\cite{ANSYS} show that increasing the distance of PCB and reducing the width of the readout strips can increase the impedance, but the narrow readout strips reduce the amount and the uniformity of the induced signal charge. As a compromise, we adopt the design of a 7-\si{mm} wide readout strip, 10-\si{mm} pitch, and 0.55-\si{mm} thick float glass.  

\begin{figure}[htbp]
\begin{minipage}[t]{0.5\linewidth}
\centering
\includegraphics*[width=1\textwidth,height=.5\textwidth]{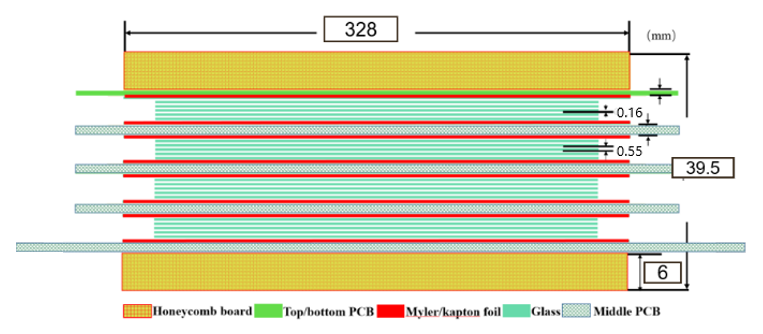}
\centerline{(a)}
\end{minipage}
\hfill
\begin{minipage}[t]{0.5\linewidth}
\centering
\includegraphics*[width=.8\textwidth,height=.5\textwidth]{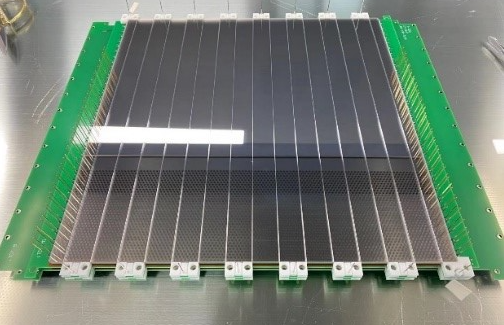}
\centerline{(b)}
\end{minipage}
\caption{Schematic representation and actual photo of 24-gap MRPC, it is divided into four stacks each.}
\label{fig:3_2}
\end{figure}

Fig~\ref{fig:3_2}(a) shows the structure of the prototype MRPC module for CEE-iTOF,which has four stacks, with six gas gaps of 0.16~\si{mm} in each stack. According to the simulation, the flux rate in the iTOF area is $\sim$50~\si{Hz/cm^2}; thus, MRPC with normal float glass can meet this rate capability. The total active area of the MRPC module is 328~\si{mm}$\times$ 303~\si{mm}, sensed by 32 readout strips. The size of each strip is 303~\si{mm}$\times$7~\si{mm}, and there is a 3-\si{mm} interval between neighboring strips. The whole detector contains five readout PCB boards. Induction signals of different polarities are read out from both ends of readout strips, then lead out to the bottom PCB through signal pins, and finally the signals are sent to FEE differentially. The thickness of glass plate we used is 0.55~\si{mm}, so that this prototype' impedance is matched to 49~\si{\Omega}, which is matched to 110~\si{\Omega} on the  bottom PCB with a resistor. \par

As a gas detector, the MRPC needs to work in a gas-tight environment. Generally, the MRPC and FEE are placed in a large gas box. On the one hand, the heat generated by FEE increases the temperature of gas, which in turn affects the performance of the MRPC and FEE. On the other hand, due to the narrow width of the gas gap, the gas exchange in gas gaps is mainly through diffusion, resulting in very low gas flow. According to the latest research results~\cite{bartos2022ageing}, low gas flow increases the MRPC aging phenomena. To solve these problems, we design a semi-sealed box structure and, use bottom PCB as a part of the gas sealing. The FEE is mounted on the surface of bottom PCB directly, located outside the gas box. The volume of the gas frame is minimized, except for some space for high voltage and gas connectors. To achieve compact structure and tight gas, we use three-dimemsional printed blocks to wrap the fishing line instead of usual nylon screws. Meanwhile, as shown in Fig~\ref{fig:3_2}(b), the PCB adopts a saw-tooth edge structure to further reduce the dead area when the MRPC modules were put together in a SM.\par

\subsection{Readout electronics}
As mentioned in section 2, the goal of the prototype MRPC is to achieve a \textless 30~\si{\pico\second} time resolution. As for readout electronics, a timing precision of less than 10~\si{ps} RMS is required. The main solutions of the readout electronics include high-speed waveform sampling and amplifier/discriminator combined with time-to-digital converter (TDC). High-speed waveform sampling is characterized by high time accuracy, which can reach several ps. However, high-speed waveform sampling chips have the characteristics of excessive power consumption,and it is difficult forthis technology to meet the actual needs of multi-channel CEE experiments, and the future high-luminosity operation in HIAF. Another technical solution uses a special chip NINO~\cite{anghinolfi2004nino} or PADI~\cite{ciobanu2014padi} to amplify and discriminate signals,measured by HPTDC~\cite{christiansen2004hptdc} or GET4~\cite{deppe2009gsi}. The NINO and PADI chip have the advantages of low power consumption and simultaneous multi-channel measurement. However, the overall time jitters of NINO/HPTDC or PADI/GET4 are usually beyond 20~\si{ps}~\cite{fan2012high}\cite{loizeau2014characterization}. The electronic team of the CEE-iTOF group achieves a high time precision measurement circuit by employing an FEE module utilizing NINO chip and a time-digitization-module(TDM) with double-chain TDL FPGA TDC, as reported in \cite{luprototype}. The time precision via a 10-\si{m}-long coaxial cable was tested using an 81180A generator in laboratory. The result indicates that the time precision varies from 7.6~\si{\pico\second} RMS to 10.5~\si{\pico\second} RMS while the charge of input signal varies from 2~\si{\pico C} to 0.05~\si{\pico C}. The charge of the MRPC signal generally exceeds 0.05~\si{\pico C}, so the overall time performance of the electronics is better than 10~\si{ps}.
\begin{figure}[htbp]
\centering
\includegraphics[width=.8\textwidth,height=.6\textwidth]{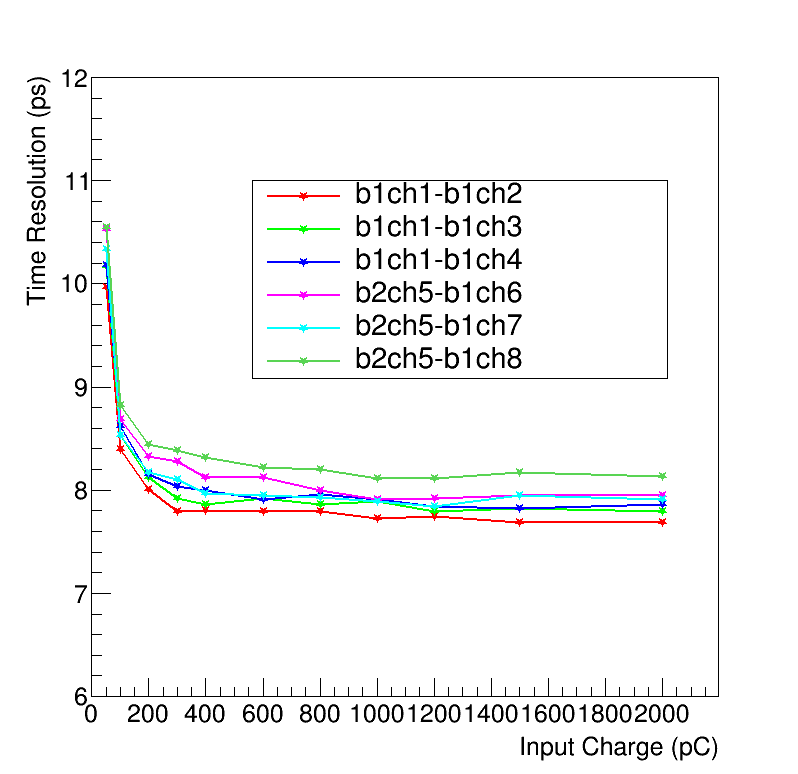}
\caption{ Time jitter of different channels varies input charge. }
\label{fig:3_3}
\end{figure}

\section{Cosmic ray test }
\subsection{cosmic ray test system }
\begin{figure}[htbp]

\begin{minipage}[t]{0.5\linewidth}
\centering
\includegraphics*[width=.9\textwidth,height=.9\textwidth]{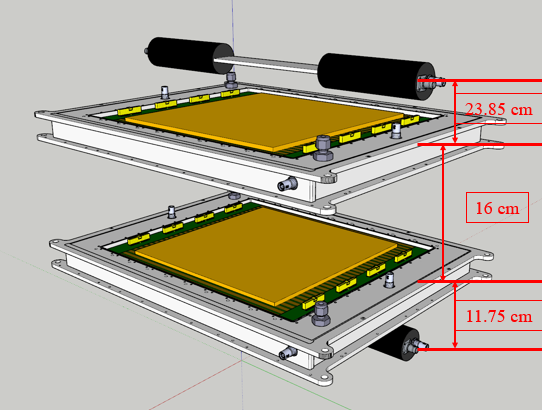}
\end{minipage}
\hfill
\begin{minipage}[t]{0.5\linewidth}
\centering
\includegraphics*[width=.9\textwidth,height=.9\textwidth]{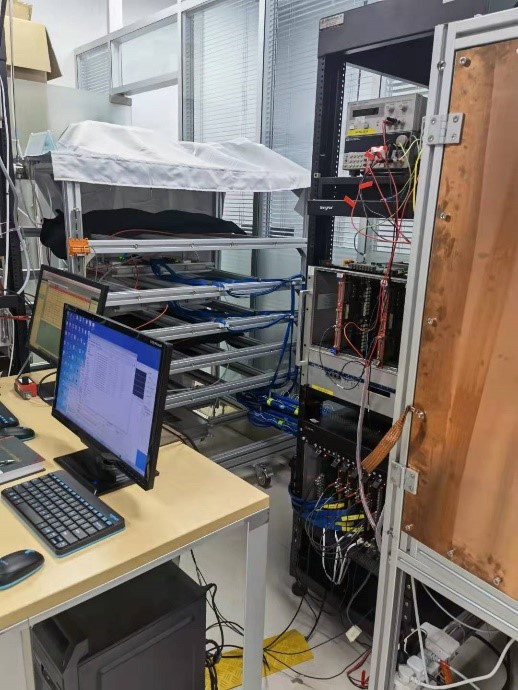}
\end{minipage}

\begin{minipage}[t]{0.5\linewidth}
\centering
\includegraphics*[width=2\textwidth,height=.7\textwidth]{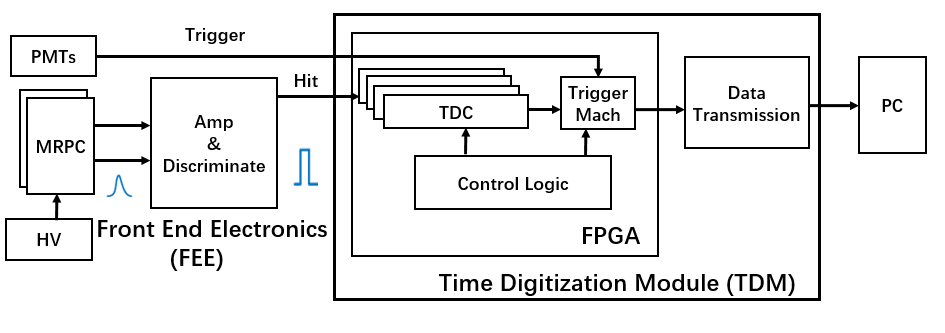}
\end{minipage}

\caption{Diagram of cosmic ray test system.  }
\label{fig:4_1}
\end{figure}

To study the performance of the prototype MRPC, a cosmic ray test platform has been set up. Fig~\ref{fig:4_1} shows the diagram of the test system. The platform comprises two scintillator detectors with four PMTs, and two identical MRPC prototypes placed in a gas-box. The working gas we used comprises 5\% iso-C4H10, 5\% SF6, and 90\% Freon. When cosmic rays passed through the two scintillators, signals from four PMTs were transmitted to a logic coincidence module(ORTEC CO4040), then the output NIM level coincidence signal was converted to a TTL level signal by CEAN N89 module,triggering the TDM module. The readout differential signals from the two MRPCs are intially amplified and discriminated by NINO-based FEE, and then the output TOT signals were transmitted vis a 10-m-high density coaxial cable (SAMTEC) to the FPGA TDC to complete the leading time and the trailing time measurements.\par

\subsection{Results of cosmic ray test }
The active area of scintillator in the cosmic ray system is 5$\times$20~\si{cm^2}. The sensitive area of the scintillator determines the effective area of the system that can be detected. The trigger scintillator detectors were placed parallel to the MRPC's readout strip. In this system, each detector has eight strips equipped with readout electronics. The detection efficiency is defined as the ratio of the counts of any of the eight strips of the MRPC signals to the counts of coincidence signal of four PMTs-$Counts_{mrpc}$ /$Counts_{PMTs}$. The MRPC in the experiment is readout from both ends of strip, so for each strip, a count with valid signals from both ends is required to be considered as a true count.\par

\begin{figure}[htbp]
\begin{minipage}[t]{0.5\linewidth}
\centering
\includegraphics*[width=1\textwidth,height=.9\textwidth]{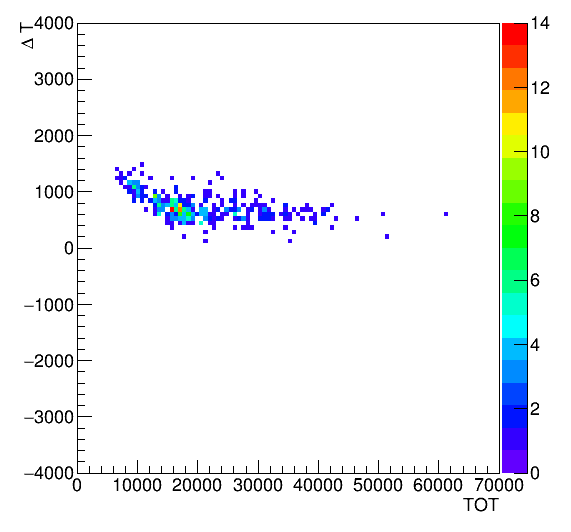}
\centerline{(a)}
\end{minipage}
\hfill
\begin{minipage}[t]{0.5\linewidth}
\centering
\includegraphics*[width=1\textwidth,height=.9\textwidth]{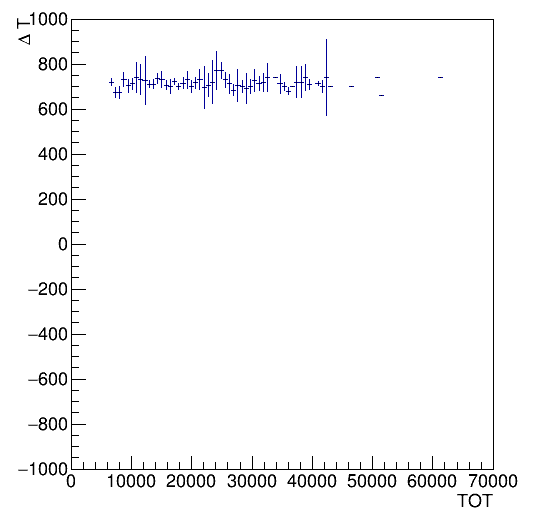}
\centerline{(b)}
\end{minipage}
\caption{$\Delta$T and TOT distribution before and after correction.}
\label{fig:4_22}
\end{figure}

In the cosmic ray system, we placed two identical MRPCs, so the time resolution of the two MRPCs can be considered identical. According to this hypothesis, two MRPCs can calibrate each other by themselves; the MRPC's time resolution can be defined as:
\begin{equation}
\sigma_{MRPC}=\frac{\sigma(T_{mrpc1}-T_{mrpc2})}{\sqrt{2}}
\label{equation:MRPC}
\end{equation}
where $T_{mrpc1}$ and $T_{mrpc2}$ are the times when the cosmic ray hits on MRPC1 and MRPC2,respectively. Based on the characteristics of the two-ended readout of the MRPC detector, $T_{mrpc1}=(T_1+T_2)/2$ and $T_{mrpc2}=(T_3+T_4)/2$, where $T_1$,$T_2$,$T_3$,and $T_4$ represent the hits time recorded by the FEE of the MRPC 1 and MRPC 2, respectively. Due to the timing nature of the readout electronics, $T_{mrpc}$ is strongly related to the pulse height, and this time walk effect needs to be corrected during data analysis. Fig~\ref{fig:4_22}(a) shows the relationship between $\Delta$T and TOT before correction($\Delta T= \frac{T_1+T_2-T_3-T_4}{2}$); Fig~\ref{fig:4_22}(b) shows the $\Delta$T-TOT relationship after the slewing correction by polynomial, and it can be seen from the figure that the time walk effect is basically corrected.\par
\begin{figure}[htbp]
\begin{minipage}[t]{0.5\linewidth}
\centering
\includegraphics*[width=1\textwidth,height=.9\textwidth]{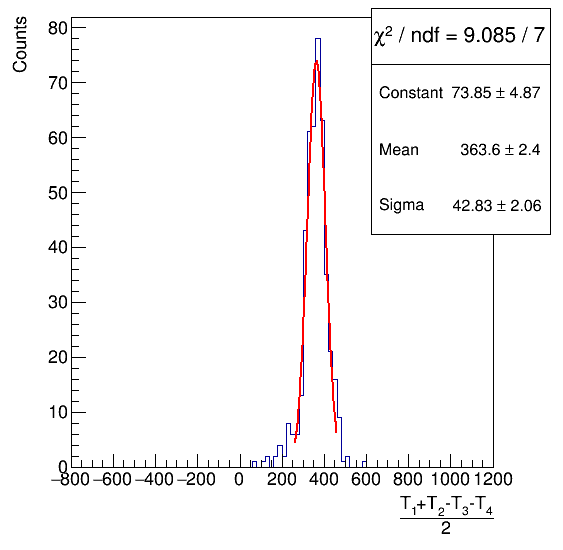}
\centerline{(a)}
\end{minipage}
\hfill
\begin{minipage}[t]{0.5\linewidth}
\centering
\includegraphics*[width=1\textwidth,height=.95\textwidth]{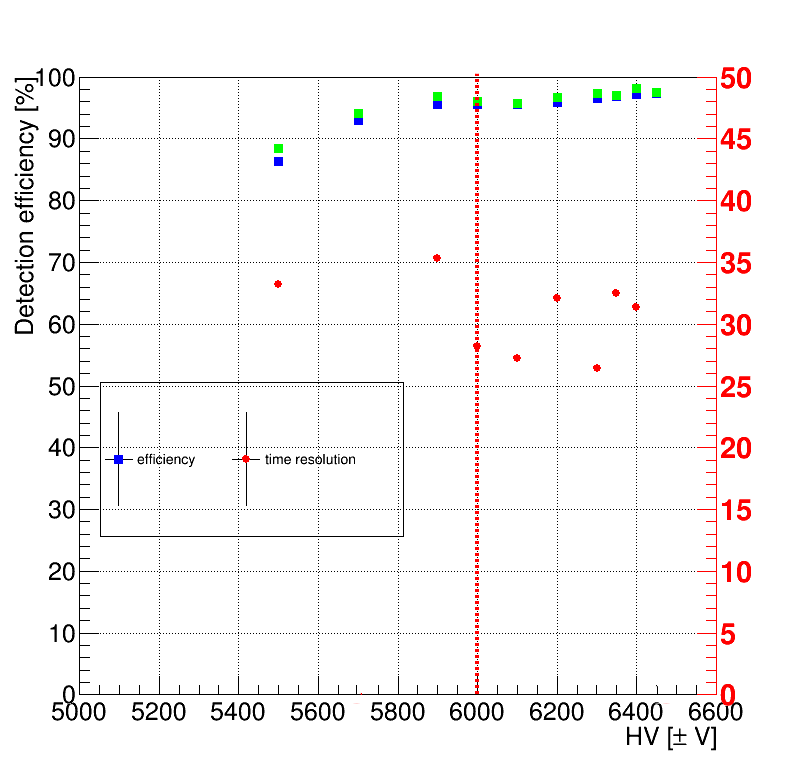}
\centerline{(b)}
\end{minipage}
\caption{Efficiency and time resolution of the MRPC prototype. }
\label{fig:4_2}
\end{figure}

In Fig~\ref{fig:4_2}(a), an example time resolution of the MRPC after correcting for the time walk has been shown. After Gaussian fitting, we get the sigma value of the distribution of 42~\si{ps}, so according to Equation~\ref{equation:MRPC}, the time resolution of the prototype MRPC is about 30 ps. 
Fig~\ref{fig:4_2}(b) shows the dependence measurement of efficiency and time resolution with high voltage. When the voltage of the MRPC is 5900~\si{V}, the efficiency enters the plateau area, which is higher than 95\%. Thus, the time resolution and detect efficiency of the prototype MRPC fulfill the CEE experiments requirement.

\subsection{Discussion }
\begin{figure}[htbp]
\centering
\includegraphics[width=.5\textwidth]{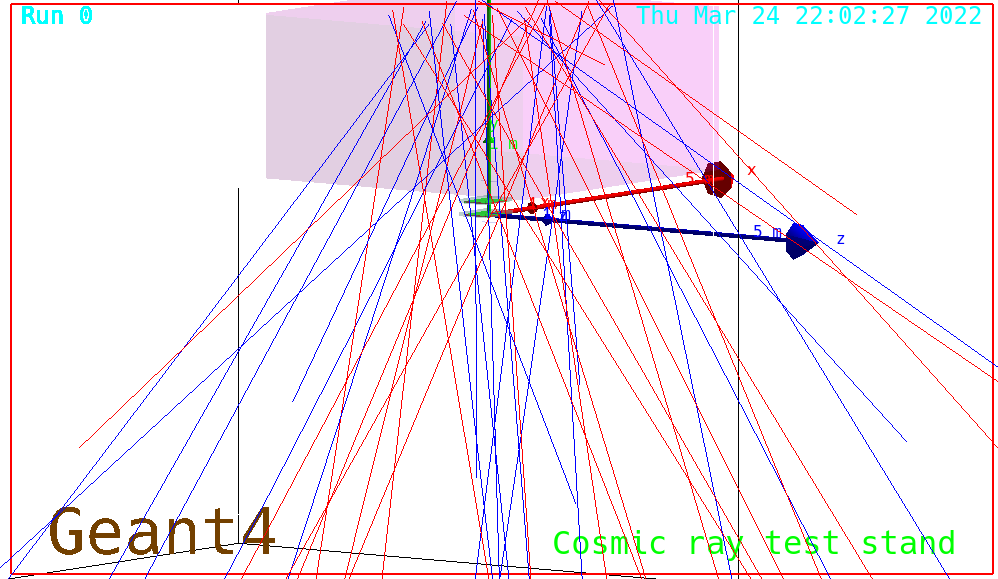}
\caption{  A typical cosmic ray process in GEANT4 simulation.  The red and blue tracks represent $\mu^{-}$ and $\mu^{+}$ respectively.}
\label{fig:5_1}
\end{figure}

\begin{figure}[htbp]
\begin{minipage}[t]{0.5\linewidth}
\centering
\includegraphics*[width=.8\textwidth,height=.9\textwidth]{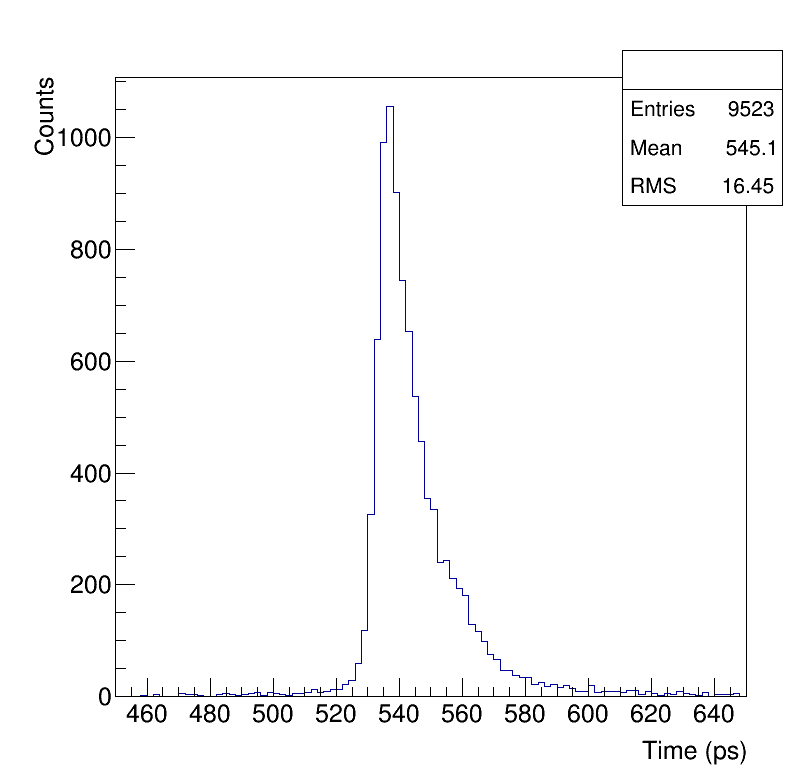}
\centerline{(a)}
\end{minipage}
\hfill
\begin{minipage}[t]{0.5\linewidth}
\centering
\includegraphics*[width=.8\textwidth,height=.9\textwidth]{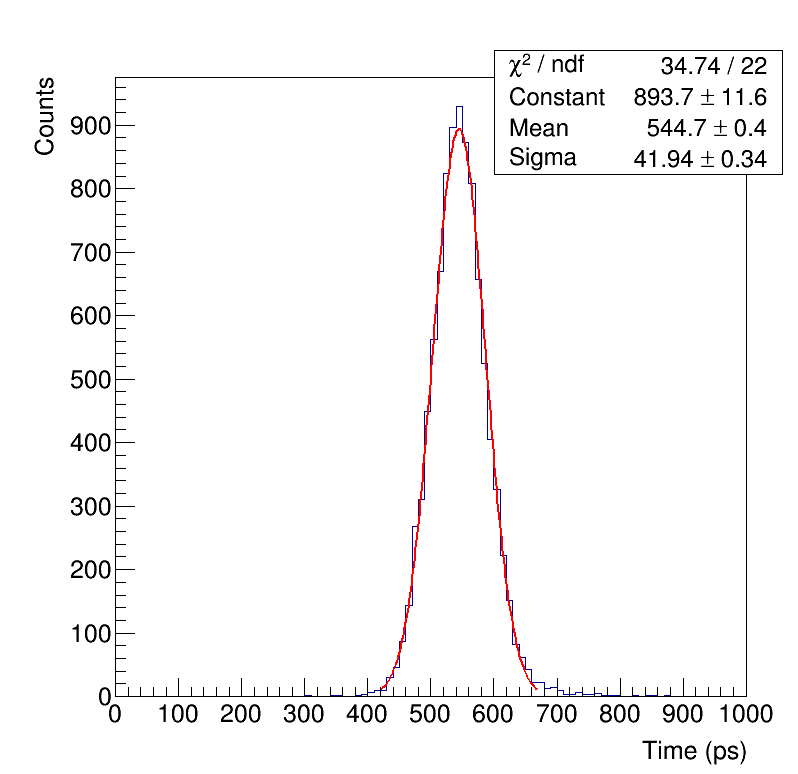}
\centerline{(b)}
\end{minipage}
\caption{Simulation results for cosmic ray test system.   }
\label{fig:5_2}
\end{figure}

In our cosmic ray test system, the distance between the two MRPCs is 16~\si{cm}. For the ultra-high time resolution MRPC detector, the time jitter of cosmic rays' momentum dispersion and flight angle should not be ignored. To correct this effect, we built a simulation frame using GEANT4 and Cosmic-ray Shower library (CRY)~\cite{hagmann2012cosmic}. The simulated geometric parameters were exactly set based on our cosmic ray test system. The typical cosmic ray process is shown in Fig~\ref{fig:5_1}. In simulation, considering the impact of the floor,cosmic rays were emitted from the surface of a 2-m-thick concrete wall , and then used two scintillators trigger select valid events. The timing of the cosmic ray passes that through MRPC1 is T1, and the timing it passes through MRPC2 is T2. T1 - T2 represents the intrinsic time jitter of the cosmic rays as shown in Fig~\ref{fig:5_2}(a). We have assumed a intrinsic time resolution series for the MRPCs in the simulation.Then,the distribution of time difference between the two MRPCs was fit using the Gaussian function.The experiments and simulations correlate best with a timing smearing of 28~\si{ps}, as shown in Fig~\ref{fig:5_2}(b). Therefore, the preliminary results indicate that the time resolution of the prototype MRPC is $~\sim$28~\si{\pico\second} by deducting  the cosmic ray's momentum and angle dispersion. 

\section{Conclusions }
The CEE experiment to be conducted at HIRFL-CSR is planned to be operational in 2025.To meet this target, all aspects of the system must be tested extensively.The PID of medium rapid region relies on iTOF system based on the MRPC technology. This study defines the performance requirements of iTOF through simulations,including the number of channels, granularity, and time resolution.To enable related technologies to be used in future HIAF, we developed a high time resolution  MRPC adopting 4$\times$6 gaps with a 160-\si{\mu m} gap size and a new readout design. A prototype with a time resolution better than 30~\si{ps} was obtained in the laboratory using a cosmic ray system. In addition, we performed a preliminary simulation of the cosmic ray system, and by comparing the simulations and experiments, the time resolution of the prototype is greater than 28~\si{\pico\second}, with detailed analysis and further testing in progress. 

\acknowledgments

The author thanks the high energy physics group of USTC. This project is supported by National Natural Science Foundation of China (U11927901). Special thank goes to Professor Qing Luo from National Synchrotron Radiation Laboratory (China) for help on FEM calculation (ANSYS).


\bibliography{wangxj}

\begin{thebibliography}{10}

\bibitem{braun2007quest}
Peter Braun-Munzinger and Johanna Stachel.
\newblock The quest for the quark--gluon plasma.
\newblock {\em Nature}, 448(7151):302--309, 2007.

\bibitem{xiao2014probing}
Zhi-Gang Xiao, Gao-Chan Yong, Lie-Wen Chen, Bao-An Li, Ming Zhang, Guo-Qing
  Xiao, and Nu~Xu.
\newblock Probing nuclear symmetry energy at high densities using pion, kaon,
  eta and photon productions in heavy-ion collisions.
\newblock {\em The European Physical Journal A}, 50(2):1--10, 2014.

\bibitem{ackermann2003star}
KH~Ackermann, N~Adams, C~Adler, Z~Ahammed, S~Ahmad, C~Allgower, J~Amonett,
  J~Amsbaugh, BD~Anderson, M~Anderson, et~al.
\newblock Star detector overview.
\newblock {\em Nuclear Instruments and Methods in Physics Research Section A:
  Accelerators, Spectrometers, Detectors and Associated Equipment},
  499(2-3):624--632, 2003.

\bibitem{aamodt2008alice}
Kenneth Aamodt, A~Abrahantes Quintana, R~Achenbach, S~Acounis, D~Adamov{\'a},
  C~Adler, M~Aggarwal, F~Agnese, G~Aglieri Rinella, Z~Ahammed, et~al.
\newblock The alice experiment at the cern lhc.
\newblock {\em Journal of Instrumentation}, 3(08):S08002, 2008.

\bibitem{friese2006cbm}
V~Friese.
\newblock The cbm experiment at gsi/fair.
\newblock {\em Nuclear Physics A}, 774:377--386, 2006.

\bibitem{toneev2007nica}
Viacheslav Toneev.
\newblock The nica/mpd project at jinr (dubna).
\newblock {\em arXiv preprint arXiv:0709.1459}, 2007.

\bibitem{xia2002heavy}
Jia-Wen Xia, Wen-Long Zhan, Bao-Wen Wei, YJ~Yuan, MT~Song, WZ~Zhang, XD~Yang,
  P~Yuan, DQ~Gao, HW~Zhao, et~al.
\newblock The heavy ion cooler-storage-ring project (hirfl-csr) at lanzhou.
\newblock {\em Nuclear Instruments and Methods in Physics Research Section A:
  Accelerators, Spectrometers, Detectors and Associated Equipment},
  488(1-2):11--25, 2002.

\bibitem{yang2013high}
JC~Yang, JW~Xia, GQ~Xiao, HS~Xu, HW~Zhao, XH~Zhou, XW~Ma, Y~He, LZ~Ma, DQ~Gao,
  et~al.
\newblock High intensity heavy ion accelerator facility (hiaf) in china.
\newblock {\em Nuclear Instruments and Methods in Physics Research Section B:
  Beam Interactions with Materials and Atoms}, 317:263--265, 2013.

\bibitem{lu2017conceptual}
LiMing L{\"u}, Han Yi, ZhiGang Xiao, Ming Shao, Song Zhang, GuoQing Xiao, and
  Nu~Xu.
\newblock Conceptual design of the hirfl-csr external-target experiment.
\newblock {\em SCIENCE CHINA Physics, Mechanics \& Astronomy}, 60(1):1--7,
  2017.

\bibitem{hu2017t0}
D~Hu, M~Shao, Y~Sun, C~Li, H~Chen, Z~Tang, Y~Zhang, J~Zhou, H~Zeng, X~Zhao,
  et~al.
\newblock A t0/trigger detector for the external target experiment at csr.
\newblock {\em Journal of Instrumentation}, 12(06):C06010, 2017.

\bibitem{li2016simulation}
He~Li, Song Zhang, Fei Lu, Chen Zhong, and Yugang Ma.
\newblock Simulation of momentum resolution of the cee-tpc in hirfl.
\newblock {\em Nuclear Techniques}, 39(7), 2016.

\bibitem{sun2018drift}
YZ~Sun, ZY~Sun, ST~Wang, LM~Duan, Y~Sun, D~Yan, SW~Tang, HR~Yang, CG~Lu, P~Ma,
  et~al.
\newblock The drift chamber array at the external target facility in hirfl-csr.
\newblock {\em Nuclear Instruments and Methods in Physics Research Section A:
  Accelerators, Spectrometers, Detectors and Associated Equipment}, 894:72--80,
  2018.

\bibitem{zhu2021prototype}
SH~Zhu, HB~Yang, H~Pei, CX~Zhao, XQ~Li, and GM~Huang.
\newblock Prototype design of readout electronics for zero degree calorimeter
  in the hirfl-csr external-target experiment.
\newblock {\em Journal of Instrumentation}, 16(08):P08014, 2021.

\bibitem{urqmd}
\url{https://urqmd.org/}.

\bibitem{lippmann2012particle}
Christian Lippmann.
\newblock Particle identification.
\newblock {\em Nuclear Instruments and Methods in Physics Research Section A:
  Accelerators, Spectrometers, Detectors and Associated Equipment},
  666:148--172, 2012.

\bibitem{agostinelli2003geant4}
Sea Agostinelli, John Allison, K~al Amako, John Apostolakis, H~Araujo, Pedro
  Arce, Makoto Asai, D~Axen, Swagato Banerjee, GJNI Barrand, et~al.
\newblock Geant4—a simulation toolkit.
\newblock {\em Nuclear instruments and methods in physics research section A:
  Accelerators, Spectrometers, Detectors and Associated Equipment},
  506(3):250--303, 2003.

\bibitem{lv2014nuclear}
M~Lv, YG~Ma, GQ~Zhang, JH~Chen, and DQ~Fang.
\newblock Nuclear modification factor in intermediate-energy heavy-ion
  collisions.
\newblock {\em Physics Letters B}, 733:105--111, 2014.

\bibitem{zichichi1987laa}
Antonino Zichichi.
\newblock The laa project (cern-ep-87-122).
\newblock {\em ICFA Instrum. Bull.}, 3(CERN-EP-87-122):17--23, 1987.

\bibitem{hu2019beam}
Dongdong Hu, X~Wang, Y~Sun, M~Shao, C~Li, H~Chen, Z~Tang, Y~Zhang, J~Zhou,
  Y~Wu, et~al.
\newblock Beam test of cbm-tof mrpc prototype.
\newblock {\em Journal of Instrumentation}, 14(09):C09014, 2019.

\bibitem{riegler2004physics}
Werner Riegler and Christian Lippmann.
\newblock The physics of resistive plate chambers.
\newblock {\em Nuclear Instruments and Methods in Physics Research Section A:
  Accelerators, Spectrometers, Detectors and Associated Equipment},
  518(1-2):86--90, 2004.

\bibitem{akindinov2004space}
AN~Akindinov, A~Alici, F~Anselmo, P~Antonioli, Y~Baek, M~Basile, G~Cara Romeo,
  L~Cifarelli, F~Cindolo, F~Cosenza, et~al.
\newblock Space charge limited avalanche growth in multigap resistive plate
  chambers.
\newblock {\em The European Physical Journal C-Particles and Fields},
  34(1):s325--s331, 2004.

\bibitem{ANSYS}
\url{https://www.ansys.com/}.

\bibitem{bartos2022ageing}
D~Bartos, C~Burducea, I~Burducea, G~Caragheorgheopol, F~Constantin, L~Craciun,
  D~Dorobantu, M~Ghena, D~Iancu, A~Marcu, et~al.
\newblock Ageing studies of multi-strip multi-gap resistive plate counters
  based on low resistivity glass electrodes in high irradiation dose.
\newblock {\em Nuclear Instruments and Methods in Physics Research Section A:
  Accelerators, Spectrometers, Detectors and Associated Equipment},
  1024:166122, 2022.

\bibitem{anghinolfi2004nino}
F~Anghinolfi, P~Jarron, AN~Martemiyanov, E~Usenko, Horst Wenninger, MCS
  Williams, and A~Zichichi.
\newblock Nino: an ultra-fast and low-power front-end amplifier/discriminator
  asic designed for the multigap resistive plate chamber.
\newblock {\em Nuclear Instruments and Methods in Physics Research Section A:
  Accelerators, Spectrometers, Detectors and Associated Equipment},
  533(1-2):183--187, 2004.

\bibitem{ciobanu2014padi}
M~Ciobanu, N~Herrmann, KD~Hildenbrand, M~Ki{\v{s}}, A~Sch{\"u}ttauf,
  H~Flemming, H~Deppe, S~L{\"o}chner, J~Fr{\"u}hauf, I~Deppner, et~al.
\newblock Padi, an ultrafast preamplifier-discriminator asic for time-of-flight
  measurements.
\newblock {\em IEEE Transactions on nuclear science}, 61(2):1015--1023, 2014.

\bibitem{christiansen2004hptdc}
Jorgen Christiansen.
\newblock Hptdc high performance time to digital converter.
\newblock 2004.

\bibitem{deppe2009gsi}
Harald Deppe and Holger Flemming.
\newblock The gsi event-driven tdc with 4 channels get4.
\newblock In {\em 2009 IEEE Nuclear Science Symposium Conference Record
  (NSS/MIC)}, pages 295--298. IEEE, 2009.

\bibitem{fan2012high}
Huanhuan Fan, Changqing Feng, Weijia Sun, Chunyan Yin, Shubin Liu, and Qi~An.
\newblock A high density time to digital converter module for besiii end-cap
  tof upgrade.
\newblock In {\em 2012 18th IEEE-NPSS Real Time Conference}, pages 1--4. IEEE,
  2012.

\bibitem{loizeau2014characterization}
PA~Loizeau, N~Herrmann, I~Deppner, C~Simon, C~Xiang, M~Ciobanu, H~Deppe,
  H~Flemming, J~Fr{\"u}hauf, M~Ki{\v{s}}, et~al.
\newblock Characterization of the get4 v1. 0 tdc asic with detector signals.
\newblock {\em GSI Scientific Report 2013}, page~40, 2014.

\bibitem{luprototype}
Jiaming Lu, Lei Zhao, Baolin Hou, Liujiang Yan, Karen Chen, Shubin Liu, and
  Qi~An.
\newblock A prototype design of the readout electronics of mrpc detectors in
  cee in hirfl.

\bibitem{hagmann2012cosmic}
Chris Hagmann, David Lange, Jerome Verbeke, and Doug Wright.
\newblock Cosmic-ray shower library (cry).
\newblock {\em Lawrence Livermore National Laboratory document UCRL-TM-229453},
  2012.

\end{thebibliography}
\bibliographystyle{unsrt}
\end{document}